\documentclass[runningheads]{llncs}
\usepackage{graphicx}
\usepackage{color}
\usepackage{algorithm}
\usepackage{algpseudocode}
\graphicspath{{./figures/}}
\usepackage{nameref}
\usepackage{hyperref}

\newcommand{\payload}{\textit{payload}}
\newcommand{\script}{\textit{script}}

\def\Let#1#2{\State #1 $\gets$ #2}
\algnewcommand{\IfThenElse}[3]{\algorithmicif\ #1\ \algorithmicthen\ #2\
\algorithmicelse\ #3}

\begin{document}
    \title{Self-Reproducing Coins as Universal Turing Machine}

\author{Alexander Chepurnoy\inst{1,2}, Vasily Kharin\inst{3}, Dmitry Meshkov\inst{1}}

\institute{Ergo Platform \\\email{catena@protonmail.com} \and
IOHK Research \\\email{alex.chepurnoy@iohk.io} \and
Research Institute\\\email{v.kharin@protonmail.com}}

    \date{\today}
    \maketitle

    \begin{abstract}
        Turing-completeness of smart contract languages in blockchain
        systems is often associated with a variety of language features
        (such as loops). 
        In opposite, we show that Turing-completeness of a blockchain system can
        be achieved through unwinding the recursive calls between
        multiple transactions and blocks instead of using a single one. We prove 
        it by constructing a simple universal Turing machine using
        a small set of language features in the unspent
        transaction output (UTXO) model, with explicitly given relations between
        input and output transaction states.
        Neither unbounded loops nor possibly infinite validation time are needed in this approach.

    \end{abstract}

    \section{Introduction}
    Blockchain technology has become widely adopted after the introduction of
    Bitcoin by S.~Nakamoto~\cite{nakamoto2008bitcoin}. This peer-to-peer
    electronic cash ledger drew the enormous attention from the public, which
    resulted in rapid development of the technology and appearance of hundreds
    of alternative cryptocurrency projects. It also turned out that the
    blockchain applications expand quite far beyond the simple ledger niche. The
    rules of transaction validation can incorporate complicated logic, which is
    the essence of so-called smart contracts. In the case of Bitcoin the logic
    is implemented in the special-purpose Script language, which is believed not
    to be Turing complete. This belief stimulated the development of other smart
    contract platforms with the emphasis on the language universality.
    Particularly, in Ethereum~\cite{buterin2014next} the $jump$ opcode was
    introduced in a virtual machine assembly language in order to incorporate
    unlimited loops.  In practice this resulted in various vulnerabilities and
    DoS attacks \cite{atzei2017survey} since transaction computation cost
    (so-called $gas$) can only be calculated in runtime.
    Moreover, Turing-completeness of Ethereum is
    still a subject of debates mostly due to the undecidability of the halting problem
    in combination with a bounded block validation time. The gas
    limit is often viewed as a fundamental component preventing
    Turing-completeness~\cite{miller2016ethereum}.

    A Turing-complete programming language is a language which allows
    description of a universal Turing machine. A universal Turing machine is
    the Turing machine which can simulate any other Turing machine; its
    existence is one of the main results of the Turing
    theory~\cite{turing1937computable}. The study of Turing machines is strongly
    motivated by the Church---Turing thesis, which states that Turing machines
    are capable of universal computation. The thesis is often viewed as a
    definition of computation and computability~\cite{turing1939systems}. The
    set of known computation devices and models was rapidly growing during the
    twentieth century, and the methods of their analysis were improved as well.
    The usual way of proving the Turing-completeness of a system,
    a device or a language is about using it to emulate a system that is already
    proven to be Turing complete. A system which we are using in this work is 
    one-dimensional cellular automaton Rule 110. It was conjectured to
    be Turing complete by S.~Wolfram~\cite{wolfram1986theory}. The conjecture
    was proven by M.~Cook~\cite{cook2004universality} based on previous
    works by E.~L.~Post~\cite{post1943formal}.

    The utter simplicity of Rule 110 makes it an appealing basis for proving Turing-completeness. 
    In the present work we construct Rule 110 automaton algorithm for UTXO blockchain
    and implement it in $\Sigma$-State smart contract language\cite{chepurnoy2017sigma}.
    We require neither loops, nor jump
    operator, nor recursive calls inside a transaction. Instead, we treat the
    computation as if it is occurring between the transactions (or maybe
    blocks). In this context transaction chaining and replication furnishes us
    with potentially infinite loops and recursion, while a combination of outputs
    for multiple transactions yields analog of a potentially infinite tape.
    The underlying idea of complexity growth is similar to the one expressed
    in~\cite{von1951general,von1966theory}.

    This paper is structured as follows: in Section~\ref{section2}, we first
    describe a naive implementation of Rule 110 using a simple Bitcoin-like
    scripting language. Then we discuss the pitfalls arising from compliance
    with the blockchain properties, and show the way to overcome them. Section~\ref{section3} 
    describes the implementation for the real-world blockchains, and also sketches the
    discussion on the nature of computation in the framework of blockchain
    scripting and validation rules.
    In \nameref{appendix1} we describe the structure of a general-purpose guarding script for an output  
    which can be transformed into an actual algorithm, along with the transformation procedure.
    
    \section{Rule 110 implementation}
    \label{section2}

    In this section we describe an implementation of Rule 110 cellular automaton. The automaton
    is transforming one-dimensional string of zeros and ones by applying evolution rules. One
    step of evolution for one bit is defined by its value $c$ together with the
    values of the two neighboring bits --- the left one $\ell$ and the right one $r$, along with a 
    transition rule defined in Algorithm~\ref{alg:calc_bit}

    \begin{algorithm}[H]
        \caption{Transition function of the Rule 110 automaton}
        \label{alg:calc_bit}
        \begin{algorithmic}[1]
            \Function{calcBit}{$\ell$, $c$, $r$}
            \State
            \Return $(\ell\wedge c\wedge r) \oplus (c\wedge r) \oplus c \oplus r$
            \EndFunction
        \end{algorithmic}
    \end{algorithm}

    For the automaton implementation in a blockchain we use Bitcoin-like
    transactions consisting of inputs and outputs. Every output consists of a
    guarding \script{} and a \payload{}, while an input is a reference to an
    output from a previous transaction.  We assume that the current state of the
    automaton is stored in the transaction output's \payload{}.  The general
    idea is to use the next transaction as a single step of the system
    evolution. In order to achieve this, two main conditions must be satisfied.
    First, the \payload{} of at least one newly generated output should contain
    the updated state of the automaton. Second, this output must contain exactly
    the same script. These conditions require the transaction input to have
    access to the output's \script{}s and \payload{}s.  It is implicitly present
    in the vast amount of existing blockchains, since in most cases scripts
    verify the signature of the spending transaction, which is constructed over
    the byte array containing the new outputs.  However, this way of accessing output's 
    data may be hardly exploitable.  In the paper we assume that the 
    guarding script of an input has direct access to the spending transaction outputs.

    Keeping all these in mind, we come to the following validation script:

    \begin{algorithm}[H]
        \caption{Script, that ensures that the transaction performs correct rule 110 transformation
        keeping the same rules for further iterations}
        \label{alg:isRule110}
        \begin{algorithmic}[1]
            \Function{validate}{in, out}
            \Function{isRule110}{inLayer, outLayer}
            \Function{procCell}{$i$}
            \Let{$\ell$}{inLayer[$i-1$ mod inLayer.size]}
            \Let{$c$}{inLayer[$i$]}
            \Let{$r$}{inLayer[$i+1$ mod inLayer.size]}
            \State
            \Return \Call{calcBit}{$\ell$, $c$, $r$}
            \EndFunction
            \State \Return outLayer = inLayer.indices.map(\textsc{procCell})
            \EndFunction
            \State
            \Return \Call{isRule110}{self.payload, out[0].payload}
            $\wedge$ (self.script = out[0].script)
            \EndFunction
        \end{algorithmic}
    \end{algorithm}

    The script performs two checks. First, it takes the payload of a current
    input and ensures, that the result of Rule 110 application equals to the
    payload of the first output. Second, it checks that the guarding script of the first output 
    is the same as a script of the input. The full implementation of this script
    in the smart contract language of an existing UTXO blockchain Ergo is
    provided at \cite{ergoScript1}.

\begin{figure}[h]
        \centering
        \includegraphics[width=.8\textwidth]{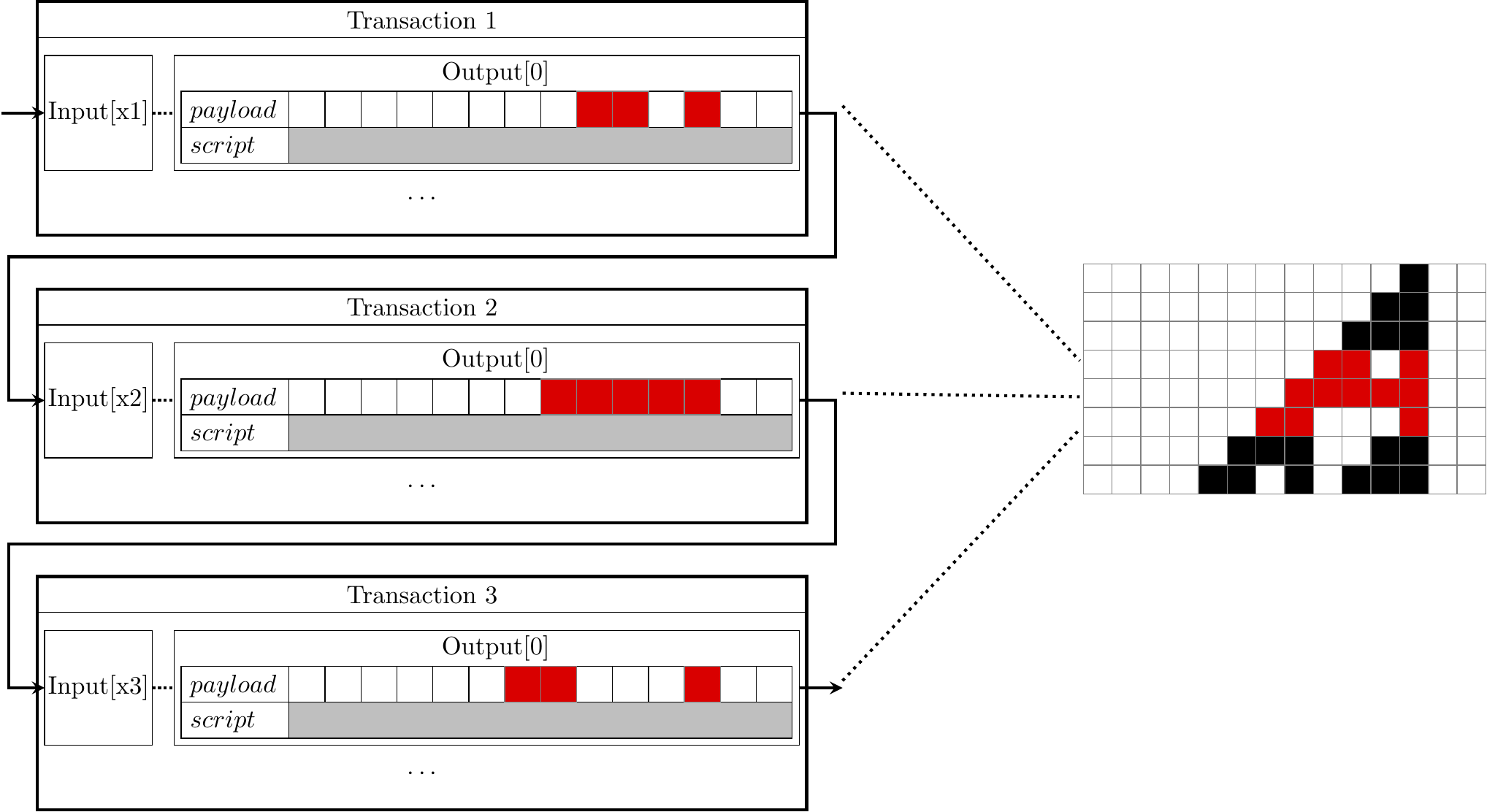}
        \caption{Transaction chain following Rule 110. See
            Alg.~\ref{alg:isRule110} for the
            $script$ field description.
        \label{fig:txs} }
    \end{figure}

    With this script, the cellular automaton evolution may be started by
    chaining transactions in a blockchain. Fig.~\ref{fig:txs} shows three
    transactions (on the left), each one representing the iteration of the
    automaton (on the right).

    Potentially infinite evolution of a cellular automaton, which is
    required for Turing-completeness, can be modeled by chaining potentially infinite
    number of transactions in the blockchain. However, there is a pitfall left.
    The size of the data stored in output must have an upper-bound, and validation
    time for a transaction must be bounded as well, otherwise blockchain is losing its security properties~\footnote{For example, in the Bitcoin backbone protocol model from~\cite{garay2015bitcoin}, block validation
    should happen within finite and a-priori known round duration.}.

\begin{figure}[h]
        \centering
        \includegraphics[width=.8\textwidth]{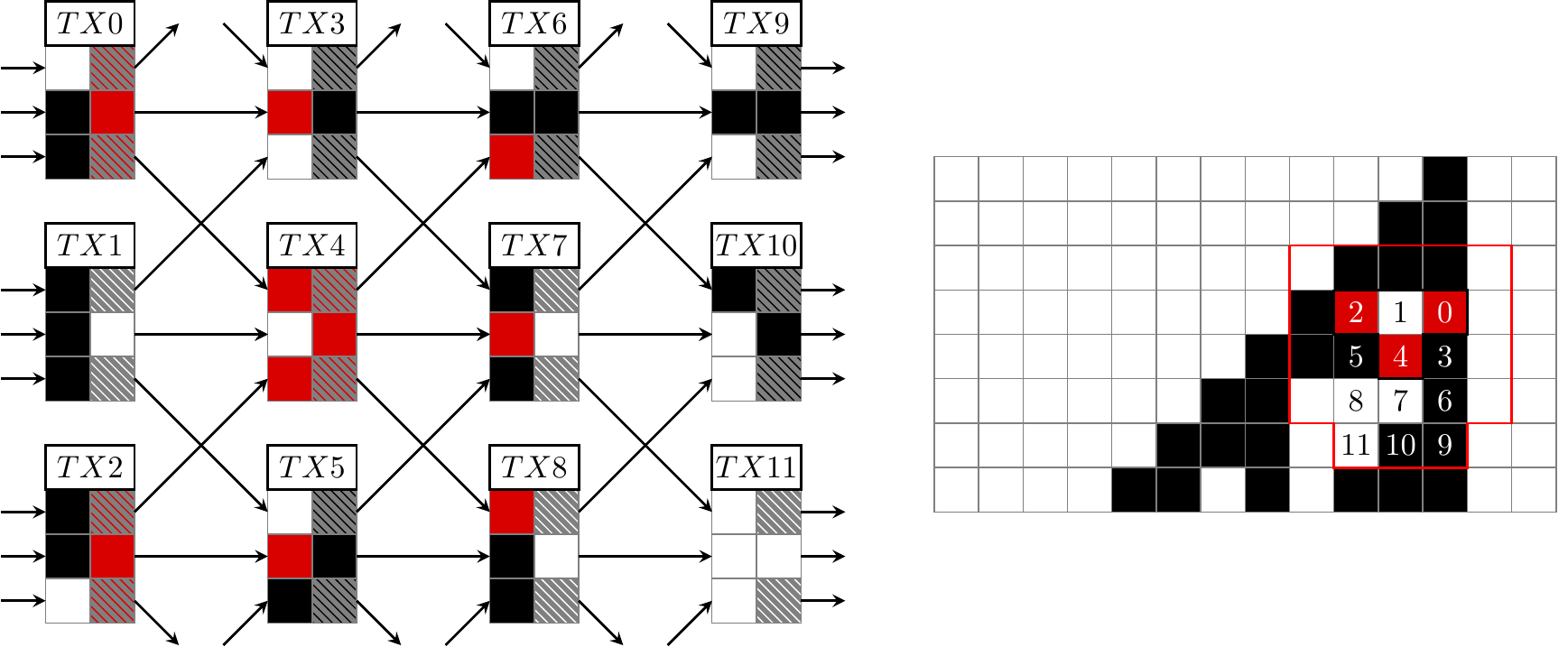}
        \caption{Evolution of the cellular automaton described in
            Alg.~\ref{alg:txBit}. Every non-boundary transaction spends three
            outputs, and generates three new ones with identical bit values.
            Hatching indicates ``mid'' flag being unset. Numbers in the cells on
            the right pane correspond to the transaction numbers on the left.
        \label{fig:bit_txs} }
    \end{figure}

    The natural workaround is to split the automaton state between
    transactions once it becomes too large. As an extreme case one can make a
    transaction output play a role of a single bit of the automaton. While being
    inefficient, this implementation keeps the logic simple and complies with the
    requirements of the blockchain and of potentially infinite evolution in
    space and time. The pseudocode of the corresponding script is
    provided in the Algorithm~\ref{alg:txBit} and its implementation in $\Sigma$-State
    contract language is provided at \cite{ergoScript2}. Fig.~\ref{fig:bit_txs}
    schematically shows the sequence of transactions (on the left), that corresponds
    to some area evaluation (on the right) of the automaton run.
    \begin{algorithm}[H]
        \caption{Validation script for the output representing the single bit, and
        the unbound grid}
        \label{alg:txBit}
        \begin{algorithmic}[1]
            \Function{verify}{in, out}\Comment{``in'' and ``out'' are lists of inputs and outputs}
            \Function{outCorrect}{out, script}
            \Comment{output structure check}
            \Let{scriptCorrect}{out[0].script = script}
            \Let{isCopy1}{out[1] = out[0].copy(mid$\leftarrow true$)}
            \Let{isCopy2}{out[2] = out[0].copy(mid$\leftarrow false$)}
            \State
            \Return ($\neg$out[0].mid) $\wedge$ scriptCorrect $\wedge$ isCopy1 $\wedge$ isCopy2
            \EndFunction
            \Function{correctPayload}{in, out}
            \Comment{output payload check}
            \State \(\triangleright\) mid flag is only set for the middle input
            \Let{inMidCorrect}{in[1].mid $\wedge$ $\neg$(in[0].mid $\vee$ in[2].mid)}
            \State \(\triangleright\) input positions are correct; n is the index of leftmost column
            \Let{inYCorrect}{(in[1].n = in[0].n) $\wedge$ (in[2].n = in[0].n)}
            \Let{inXCorrect}{(in[1].x = in[0].x$+1$) $\wedge$ (in[2].x = in[1].x$+1$)}
            \State \(\triangleright\) bits satisfy Rule 110
            \Let{inValCorrect}{out[0].val=\Call{calcBit}{in[0].val, in[1].val, in[2].val}}
            \State \(\triangleright\) output position matches the input one
            \Let{outPosCorrect}{out[0].x = in[1].x $\wedge$ (out[0].n = in[0].n$-1$)}
            \State
            \Return inValCorrect  $\wedge$ inXCorrect $\wedge$ inYCorrect $\wedge$
            \par  \hskip\algorithmicindent \hskip\algorithmicindent
            inMidCorrect $\wedge$ outPosCorrect $\wedge$ in.size=out.size=3
            \EndFunction
            \If{in[0].x=in[0].n $\wedge$ in.size=1}
            \Comment{leftmost --- add 2 zeros to the left}
            \Let{middle}{in[0].copy(x$\leftarrow$in[0].n$-1$, val$\leftarrow$0, mid$\leftarrow true$)}
            \Let{left}{in[0].copy(x$\leftarrow$in[0].n$-2$, val$\leftarrow$0, mid$\leftarrow false$)}
            \Let{realIn}{left ++ middle ++ in}
            \ElsIf{in[0].x=in[0].n $\wedge$ in.size=2}
            \Comment{next to leftmost --- add 0 to the left}
            \Let{left}{in[0].copy(x$\leftarrow$in[0].n$-1$, val$\leftarrow$0, mid$\leftarrow false$)}
            \Let{realIn}{left ++ in}
            \ElsIf{in[0].x=$-1$ $\wedge$ in.size=2}
            \Comment{rightmost --- add 0 to the right}
            \Let{right}{in[0].copy(x$\leftarrow 1$, val$\leftarrow$0, mid$\leftarrow false$)}
            \Let{realIn}{in ++ right}
            \Else
            \Comment{normal cell}
            \Let{realIn}{in}
            \EndIf
            \State
            \Return \Call{correctPayload}{realIn, out} $\wedge$ \Call{outCorrect}{out, in[0].script}
            \EndFunction
        \end{algorithmic}
    \end{algorithm}

	The script works as follows. Every output's payload contains its bit value
    $val$, 
    the column index $x$, and the minimal $x$ index at the current step $n$.
    As the
    grid expands by one at every step, $-n$ also serves as the row number. By
    default, the transaction spends three inputs (corresponding to the three
    neighboring bits from the previous row), and creates three outputs with the same
    bit value by the automaton rule. One output flagged by $mid$ is supposed to be
    spent for new value with the column number $x$, and another two --- for the
    columns $x\pm 1$ (see Fig.~\ref{fig:bit_txs}). In case the transaction creates the
    boundary cells, either one or two inputs are emulated to have zero bit values
    (lines 20--32).  The overall validation script checks the correctness of the
    positions of inputs (lines 12 and 13) and outputs (line 17), correspondence of
    of the bit values (line 15), the correctness of the $mid$  flag assignment for
    inputs (line 10) and the fact that all outputs are identical except the $mid$
    flag, which is set only once (lines 2--7).

    Since the Turing-completeness of Rule 110 was proven
    in~\cite{cook2004universality}, we conclude that even though the scripting
    language itself does not allow loops, Turing-completeness of the
    system can be achieved by combining multiple transactions together. Note
    that our language requirements are not very demanding, just about bit operations, comparisons, assignments, 
    and by-index access.

    \section{Discussion}
    \label{section3}

    The crucial move in our work is unwinding recursive calls by means
    of transaction chaining, although the language we use contains neither cycles nor
    recursion. By doing this we let a program to be executed over a sequence of
    transactions and blocks. This approach allows us to run programs in potentially infinite time on top 
    of the blockchain while there is a strict upper-bound for block validation time. 

    A single transaction in the blockchain approximately corresponds to a single
    step of a computing machine.
    The step may be as complex as language built-ins allow; however, for
    security reasons it should be possible to estimate its running time before the actual evaluation.

    One can wonder how evolving data structures~(a blockchain and a
    corresponding UTXO set) along with programmable validation rules constitute
    a Turing machine. Obviously, we do need to include clients, forming transactions,
    and honest majority of miners, including transactions into blocks, as a
    component of the machine as well --- their efforts are making the input tape of
    the machine. The same is true for Ethereum and other blockchains with smart
    contracts: the blockchain as a data structure does not endorse any
    computations --- they should be initialized by a client.



    Our approach can be used for Turing-completeness
    proofs of various smart contract languages in general. For example, it might be
    possible to prove
    that smart contracts of Waves platform \cite{wavesSmarts} are actually
    Turing complete, although the authors claimed the opposite.
    Rule 110 implementation is not required for practical cases, it just
    guarantees that any algorithm can be potentially implemented.
    Despite existence of this guarantee, efficient usage of
    self-reproducing coins in practice may require new machinery, including development
    environments and high-level smart contract languages for the multiple-transactions computations.




    \section*{Acknowledgments}

    Authors thank Manuel Chakravarty, Oksana Klimenko, and Georgy Meshkov for
    the discussions and helpful comments on early drafts of this paper.

    \bibliography{sources}

    \section*{Appendix}
    \label{appendix1}

    This section addresses a question of guarding script conversion into
    the procedure being executed by a client or a miner. Note that the guarding
    script itself does not explicitly prescribe the course of computational
    actions needed to produce a valid transaction. It rather describes the
    algorithm of telling whether the result of the actions is correct or not. As an
    example, one could set a guarding script in the form
    $5^{out[0].x}\,\textrm{mod}\, 23 = 13$. This script structure is admissible,
    but it is hard to say that it describes an actual program of discrete
    logarithm calculation. In our particular case the solution is simple. If the
    guarding script is of the form $(out[0].x=f(in))\wedge(something)$ with $f$
    being some function, then in order to satisfy the condition one can replace
    the equality check with a variable assignment. Hence if we require the script
    to be conjunction of equality checks containing the fields of the outputs
    solely on the left hand sides, and functions of the inputs on the right hand
    sides, then it actually defines the program (assuming that the inputs are fixed). It
    is fully present in the Alg.~\ref{alg:isRule110}. Another problem is collecting the right set
    of inputs for the transaction. Suppose one wants to spend $in[0]$. If the
    condition for $in[1]$ is conjunction of the expressions of type
    $in[1].x=f(in[0])$, then finding the suitable $in[1]$ is the lookup over the
    possible inputs with field $x$ being the key. Therefore, if the guarding
    script can be represented in the form
    \begin{eqnarray}
        \label{eq:scr}
        \nonumber
        &\left(\bigwedge_i\bigwedge_j(out[i].x_j=f_{ij}(in))\right)\wedge \\ 
        &\left(\bigwedge_i in[1].x_i  = g_{1i}(in[0])\right)\wedge
        \left(\bigwedge_i in[2].x_i =
        g_{2i}(in[0],in[1])\right)\wedge\dots,
        \label{eq:convertable}
    \end{eqnarray}
    it can be efficiently converted to the transaction generation algorithm:

    \begin{algorithm}[H]
        \caption{Transaction creation algorithm}
        \label{alg:tx}
        \begin{algorithmic}[1]
            \For{$in[0]~\gets~UTXO$}
                \Let{i}{0}
                \While{$scripts~of~in[0]...in[i]~have~rule~g()~for~in[i+1]$}
                    \Let{in[i+1]}{UTXO(g(in[0]...in[i]))}
                    \Let{i}{i+1}
                \EndWhile
                \Let{j}{0}
                \While{$scripts~of~in[0]...in[i]~have~rule~f()~for~out[j]$}
                    \Let{out[j]}{f(in[0]...in[i])}
                    \Let{j}{j+1}
                \EndWhile
                \If{tx(in,out).isValid}
                    \Return tx
                \EndIf
            \EndFor
        \end{algorithmic}
    \end{algorithm}
    Here the last if-statement is the consistency check. Note that both
    Alg.~\ref{alg:isRule110} and~\ref{alg:txBit} can be represented as the
    desired form~(\ref{eq:convertable}) with the length checks.

\end{document}